# Nano-Cross-Junction Effect on Phonon Transport in Silicon-Nanowire-Cages

Dengke Ma (马登科)[1,2,#], Hongru Ding (丁鸿儒)[1,2,#], Han Meng (孟涵)[1,2], Lei Feng[3], Yue Wu[4,5], Junichiro Shiomi[3,6,*], and Nuo Yang (杨诺)[1,2,*]

[1]State Key Laboratory of Coal Combustion, Huazhong University of Science and Technology (HUST), Wuhan 430074, P. R. China

[2]Nano Interface Center for Energy(NICE), School of Energy and Power Engineering, Huazhong University of Science and Technology (HUST), Wuhan 430074, P. R. China

[3]Department of Mechanical Engineering, The University of Tokyo, 7-3-1 Hongo, Bunkyo, Tokyo 113-8656, Japan

[4]School of Chemical and Environmental Engineering, Shanghai Institute of Technology, Shanghai 200235, P.R. China

[5]Department of Chemical and Biological Engineering, Iowa State University, 2033 Sweeney Hall, Ames, IA, 50011, United States

[6]Center for Materials research by Information Integration, National Institute for Materials Science, 1-2-1 Sengen, Tsukuba, Ibaraki 305-0047, Japan

[#]D. M. and H. D. contributed equally to this work.

Electronic mail: N.Y. (nuo@hust.edu.cn) and J.S. (shiomi@photon.t.u-tokyo.ac.jp)

# ABSTRACT

Wave effects of phonons can give rise to controllability of heat conduction beyond that by particle scattering at surfaces and interfaces. In this work, we propose a new class of 3D nanostructure: a silicon-nanowire-cage (SiNWC) structure consisting of silicon nanowires (SiNWs) connected by nano-cross-junctions (NCJs). We perform equilibrium molecular dynamics (MD) simulations, and find an ultralow value of thermal conductivity of SiNWC, 0.173 $Wm^{-1}K^{-1}$, which is one order lower than that of SiNWs. By further modal analysis and atomistic Green's function calculations, we identify that the large reduction is due to significant phonon localization induced by the phonon local resonance and hybridization at the junction part in a wide range of phonon modes. This localization effect does not require the cage to be periodic, unlike the phononic crystals, and can be realized in structures that are easier to synthesize, for instance in a form of randomly oriented SiNWs network.

Over the past decades, nanostructures have attracted great attentions due to its unique properties, including the low thermal conductivity. Most-commonly exercised approach is to lower thermal conductivity by phonon scattering at boundaries (surfaces and interfaces) that becomes dominant over intrinsic scattering as the length scales of the nanostructures decreases. Taking silicon nanowires (SiNWs) as a representative material, reduction of thermal conductivity has been realized by enhanced phonon scatterings at surfaces or boundaries due to high surface-to-volume ratio.[1,2] A number of works followed to further reduce the thermal conductivity of SiNWs by surface roughness, [3-5], inner holes,[6,7] dopants,[8] and kinks. [9,10] This includes report of ultra-low value 1.1 $Wm^{-1}K^{-1}$ in experiment[7] and 0.40 $Wm^{-1}K^{-1}$ in simulation[8], however, they are obtained for ultra-short nanowires of just a few nanometers. Such materials are not scalable by itself because the thermal conductivity would increase with the length,[11] and fundamental barriers remain against the practical implementation.

Another line of effort to further reduce thermal conductivity which works on bulk materials is to utilize wave nature of phonons. Periodic phononic crystals can terminate or inhibit phonon propagation by inducing interference of phonons reflected at boundaries [12-15]. A challenge here is to ensure the occurrence of wave interferences, which requires strict periodicity of the internal structure with a size on the order of the phonon waves, which is about 1 nm at room temperature.[16] In addition, boundaries of the internal structures need to be smooth enough to specularly reflect phonons. These make production of the phononic crystals by bottom-up synthesis and top-down fabrication extremely challenging.[17]

Therefore, there is a strong need for a structure that can give rise to wave effects (interference, localization, and resonance) "locally" so that the periodicity is no longer required. Recently, by introducing small pillars on a silicon thin film, Davis et al. theoretically obtained the reduction of the thermal conductivity through the local resonance of phonons.[18] In the same year, Haoxue et al. confirms local phonon interference by germanium atoms embedded at a Si interface.[19] However, these work also show limitation in the range phonon mode that the local structure can induce distinct resonance and hybridization. To have stronger impact on the overall thermal conductivity, the structure needs to influence a broad range of phonon modes instead of resonating with specific modes.

The current work aims to resolve the above issues being encouraged by the advance in the bottom-up synthesis of various nanowire structures, which has been a driving force to realize entirely new device concepts and functional system.[20] The advances include the growth of branched and hyperbranched nanowire structures[21,22], two-nanowire-heterostructures-based nanocomposites [23,24], ZnO nanotetrapod bridging networks,[25] and planar nanowire cross-junction architectures.[26] These works have shown the advantages of two-dimensional cross-junction over "bridge" junction.

In this letter, based on the above bottom-up approach and planar nano-cross-junctions (NCJs),[26] we take a step further and propose a silicon-nanowire-cage (SiNWC) (Fig. 1(c)) structure consisting of SiNWs (Fig. 1(a)) and 3D-NCJs (Fig. 1(b)). Thus, the 1D SiNW is turned into a 3D bulk material as shown in Fig. 1(c). We perform equilibrium MD simulations and find that the thermal conductivity of SiNWC ($\kappa_{NWC}$) is drastically lower than that of the SiNWs. In addition to the parameter studies by varying the cross section area (CSA) of cage bar, period length (PL), temperature, the mechanism of the reduction is identified by modal analysis and atomistic Green's function (AGF) calculations. Finally, the comparative study of SiNWC with and without periodicity confirms the localness of the resonance and hybridization effects and thus the robustness to loss of periodicity and ordered structure, leading us to suggest practical realization in a form of random networks.

Figure 1(c) shows the simulation cell in MD. The periodic boundary condition is applied in all three directions in simulations. (Simulation details are given in Appendix I) Upon studying the $\kappa_{NWC}$, (The volume is the solid part of SiNWC excluding the void) the dependence of the value on the size of simulation cell was checked, and it was confirmed that the size of 9.78×9.78×9.78 nm$^3$ adopted throughout this study was large enough to overcome the finite size effect (shown in Fig. A2).

We calculated the dependence of $\kappa_{NWC}$ on CSA of cage bar at room temperature. While, the PL of simulation cell is fixed as 4.89 nm. As shown in Fig. 2(a) (black dots), when the CSA is decreased from 7.37 nm$^2$ to 0.29 nm$^2$, the $\kappa_{NWC}$ is sharply reduced from 1.93 Wm$^{-1}$K$^{-1}$ to 0.173 Wm$^{-1}$K$^{-1}$, which is caused not only by the increasing surface-to-volume ratio, but more interesting, by the nano-cross-junction effect (details explained later and shown in Fig. 3 and 4). We also calculated the $\kappa_{NWC}$ for several different PLs at room temperature, while the CSA of cage bar is fixed as 1.18 nm$^2$. As shown in Fig. 2(a) (red squares), when reducing PL from 14.12 nm to 4.89 nm, $\kappa_{NWC}$ decreases from 0.80 Wm$^{-1}$K$^{-1}$ to 0.49 Wm$^{-1}$K$^{-1}$. One reason is that the reducing PL increases the density of NCJs, and NCJs gives rise to phonon localization (details explained later and shown in Fig. 3 and 4). Another reason is that small PL corresponds to short cage bar. Seeing the cage as a SiNW, as shown in previous studies, its thermal conductivity increases with the length.[11] Therefore, the shortening of the cage bar also decreases $\kappa_{NWC}$. (For the convenience of comparing with the measurement results, the thermal conductivity calculated by adopting in Eq. (A2) the entire volume is given in Appendix I)

To show that our results are robust, we also have used Tersoff potential [27,28] to calculate the thermal conductivity of SiNWC (as shown in Fig. 2(a)). The thermal conductivity calculated by Tersoff potential is almost the same with Stillinger-Weber potential. A more detailed calculation shows that the difference is within 10%.

The temperature dependence of $\kappa_{NWC}$ from 300 to 1000K are shown in Fig. 2(b). The structures of SiNWC calculated have different PL, as 4.89 nm, 8.15 nm and 11.9 nm. The results show that the $\kappa_{NWC}$ of SiNWC is insensitive to the temperature. A similar temperature dependence is also found in other nanostructured materials with low thermal conductivity, like kinked SiNWs[10] and 3D Si PnCs[12]. Noting that the thermal conductivity of bulk Si is dominated by Umklapp phonon-phonon scatterings, and decrease as $\sim T^{-1}$ at high temperature, the temperature independence is a signature that SiNWC thermal conductivity is dominated by surface scattering or/and the phonon localization induced by NCJs.

To show the effect of NCJ clearly, we compare the $\kappa_{NWC}$ with the thermal conductivity of SiNWs whose length and cross section area are the same as the PL of SiNWC and the CSA of cage bar, respectively, named as SiNWs.S1. The cross section area of SiNWs[10] is 1.21 nm$^2$ which is almost the same as that of cage bar, 1.18 nm$^2$. However, the length is larger than the PL of SiNWC. As the thermal conductivity of SiNWs increases linearly with the length when the length is much smaller than mean free path [11]. So, we estimated the thermal conductivity of SiNWs.S1 by scaling down the values of SiNWs from Jiang et al.[10] linearly (results shown in Fig. 2(b)). Interestingly, the thermal conductivity of SiNWC is still around one order of magnitude lower than that of SiNWs.S1. Considering the size of SiNWs.S1 is the same as the cage bar, the further reduction of the $\kappa_{NWC}$ should come from the NCJ effect, instead of surface scatterings and length confinements. As shown in Fig. A5, the similar relaxation time between SiNWs and SiNWC further ensures this point. [4,18,29] (calculation details are given in SI IV)

The obtained $\kappa_{NWC}$ is around three orders of magnitude lower than that of bulk Si calculated by EMD. The $\kappa_{NWC}$ is around 4% of that of SiNWs as reported from Yang et al.[30] and Majumdar et al.[31] What's more, the thermal conductivity of SiNWC is even lower than that of kinked SiNWs[10] (~2.6 Wm$^{-1}$K$^{-1}$) and that of Si PnCs[12] (~0.22 Wm$^{-1}$K$^{-1}$), which means the NCJ effect is stronger than the pinching effect in kinked SiNWs and periodic SiPnCs with confined heat transfer. The $\kappa_{NWC}$ is also lower than the measured thermal conductivity in superlattice Si/Ge NWs (~5.8 Wm$^{-1}$K$^{-1}$) where has intense alloy scatterings[32]. Therefore, we propose a nanostructured bulk material, SiNWC, whose thermal conductivity can be drastically lower than the nanowires.

To understand the underlying physical mechanism and explicitly show the NCJ effect, we carried out a vibrational eigen-mode analysis of phonons in SiNWs and SiNWC.[33,34] As seen in the phonon dispersion relations of SiNWs (Fig. A4(a) and Fig.A4(c)) and SiNWC (Fig. 3(a), Fig. A4(b) and Fig. A4(d)), there are flat bands across the entire frequency range in SiNWC (results calculated by Tersoff potential also shown in Fig. A4 ), which is the signature of local resonance.[18] These resonance modes interact with the propagating modes and form a hybridization, which will reduce their group velocity and hinder the transport of these propagating modes and result in phonon localization.[18] In comparison with the pillared thin film of Davis et al. [18], SiNWC exhibits more flat bands because cross junction of SiNWs introduces more resonance due to pillars extended multiple directions with larger junction area realizing more intensive mode hybridization.(A more detailed quantitative comparison is given in Appendix X)

In Fig. 3(b), by using the AGF calculation,[35,36] we get the phonon transmission coefficient of single NCJ, and compare to SiNWs with the same length (calculation details are given in Appendix VI ). The transmission coefficient of single NCJ (the red line) is much lower than that of SiNWs (the black line) in a wide range of frequency. This implies that many phonons can't transport from the one side to the other side when the NCJ exits, which is an evidence for phonon localization. (A localization analysis of the SiNWs and SiNWC by participation ratio is given in Appendix VII and VIII)

To observe the spatial distribution of localizations, we calculate the spatial energy distribution for strong localized eigen-modes. [6,33](calculation details are given in Appendix VII) Fig. 4(a) and 4(b) show the normalized energy distribution of almost localized modes (participation ratio < 0.2) in a SiNWs and a SiNWC, respectively. As shown in Fig. 4(a), in case of, the energy is likely to be localized on the surface, and the distribution is homogeneous along the longitudinal direction. Therefore, there is less localizations inside the SiNWs, opening a channel for phonons to propagate from one side to the other in the longitudinal direction. On the other hand, Fig. 4(b) shows that, in case of SiNWC, the energy is localized not only on the surface, but at the NCJ where the values of $E_i$ are much higher than those on surface, which inhibits the propagation of phonons. Better view of the localization at NCJ atoms can be obtained in Fig. 4(c), where we extract only the cage bar from Fig. 4(b). More interestingly, the NCJ atoms have a higher localization than the surface atoms. Hence, while phonons transport mainly inside NW/cage bar, the high localization at the NCJ effectively act as the bottleneck of thermal transport. (The spatial distribution of localizations calculated by using the Tersoff potential is given in Appendix IX)

As the localization is induced by local resonance, and occurs in the junction part (as shown in Fig. 4). So nano-cross-junction effect does not require with the periodic structures, which makes the SiNWC much easier to fabricate in experiment. To demonstrate this, we construct another structure, which has the same supercell length and cross section area, but a random supercell structure (shown in Fig. 1(d) and Fig. A3). The thermal conductivity of this random structure is 0.51 $Wm^{-1}K^{-1}$ (shown in Fig. 2(a)). Comparing with the thermal conductivity of the corresponding periodic structure, 0.49 $Wm^{-1}K^{-1}$, the deviation is only 4%. What's more, we construct another random structure which is consist of both 2D-NCJ and 3D-NCJ (shown in Fig. A3). Its thermal conductivity is 0.14 $Wm^{-1}K^{-1}$, which is much lower than that of the period structure whose thermal conductivity is 0.49 $Wm^{-1}K^{-1}$ (shown in Fig. 2(a)). This further reduction is due to the increasing of NCJ numbers. These two results of the random structure confirm the robustness to the loss of periodicity. What's more, as shown in Fig. 3(b), the transmission coefficient of single NCJ is much lower than that of SiNWs in a wide range of frequency, which means that the nano-cross-junction effect works well in single NCJ. The two NCJ structure (3-dimensional atomistic presentations of these structures as shown in Fig. A6), one is symmetric (the red line) and the other is random (the blue line), have almost the same transmission coefficient. These further demonstrated that the localization effect induced by nano-cross-junction does not require the cage to be periodic.

In summary, based on the nano-cross-junction, the new SiNWC structure we proposed can turn the 1D SiNWs into a 3D bulk material. By performing equilibrium MD simulations, it is found that the SiNWC has an ultralow thermal conductivity (lowest obtained as 0.173 $Wm^{-1}K^{-1}$), even compared with the extremely short SiNWs with the same cross section area. It is also found that the thermal conductivity of SiNWC increases with the increasing of the PL and the CSA of cage bar. Moreover, we found that the thermal conductivity was not sensitive to temperature on the range from 300 K to 1000 K. After a comparatively studying on the phonon eigen-modes in SiNWC and SiNWs by lattice dynamics, we demonstrated that the new type phonon localization is induced by phonon local resonance and hybridization at the junction part. As the mechanism is local effect, so it is completely different from the traditional periodic phononic crystal.[12,14] Through the MD and AGF calculation, we further demonstrated that this localization effect does not require the cage to be periodic. The novel random nano-junction structures will be much easier to fabricate than the periodic structures. That is, our work achieves a substantially advances in how to synthesis new bulk structures with ultra-low thermal conductivity.

## Acknowledgments

N. Y., D. M., H.D., and H.M. are supported by the National Natural Science Foundation of China (51576076). J. S. and L. F. are supported by JSPS KAKENHI Grant Number 16H04274. Y. W. thanks the support from The Eastern Scholar Program. The authors are grateful to Takuma Shiga, Xin Qian and Masato Onishi for useful discussions. The authors thank the National Supercomputing Center in Tianjin (NSCC-TJ) for providing help in computations.

**Competing financial interests**: The authors declare no competing financial interests.

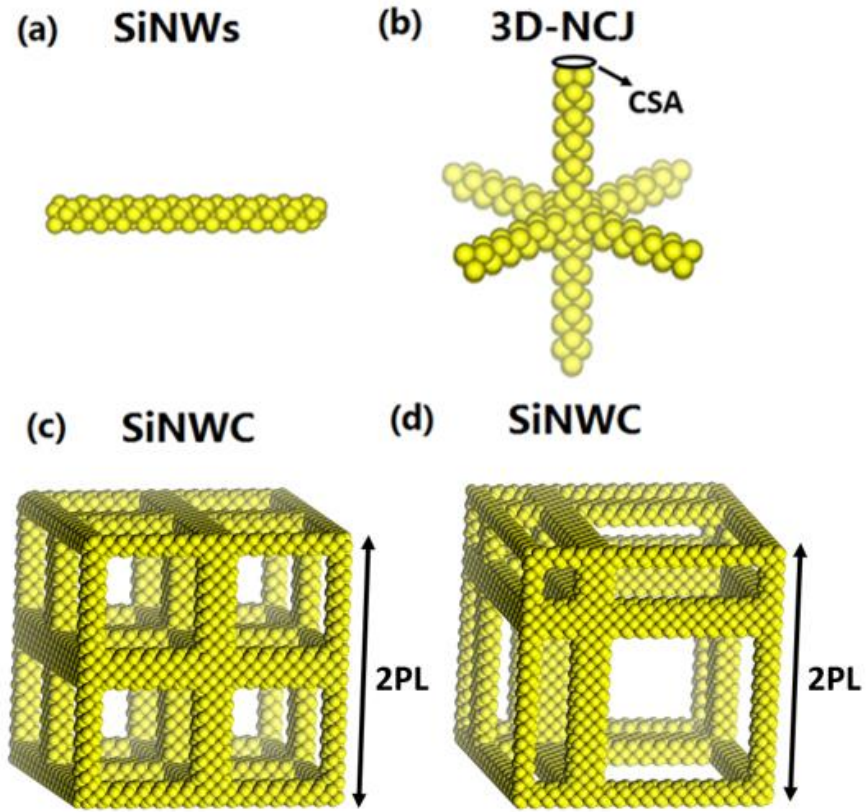

Figure 1. The bulk structure, (c) silicon nanowire cage (SiNWC) is constructed by using the 1D (a) SiNWs and (b) the 3D nano-cross-junction (NCJ). (d) A random supercell structure which has the same supercell length and cross section area as supercell (c).

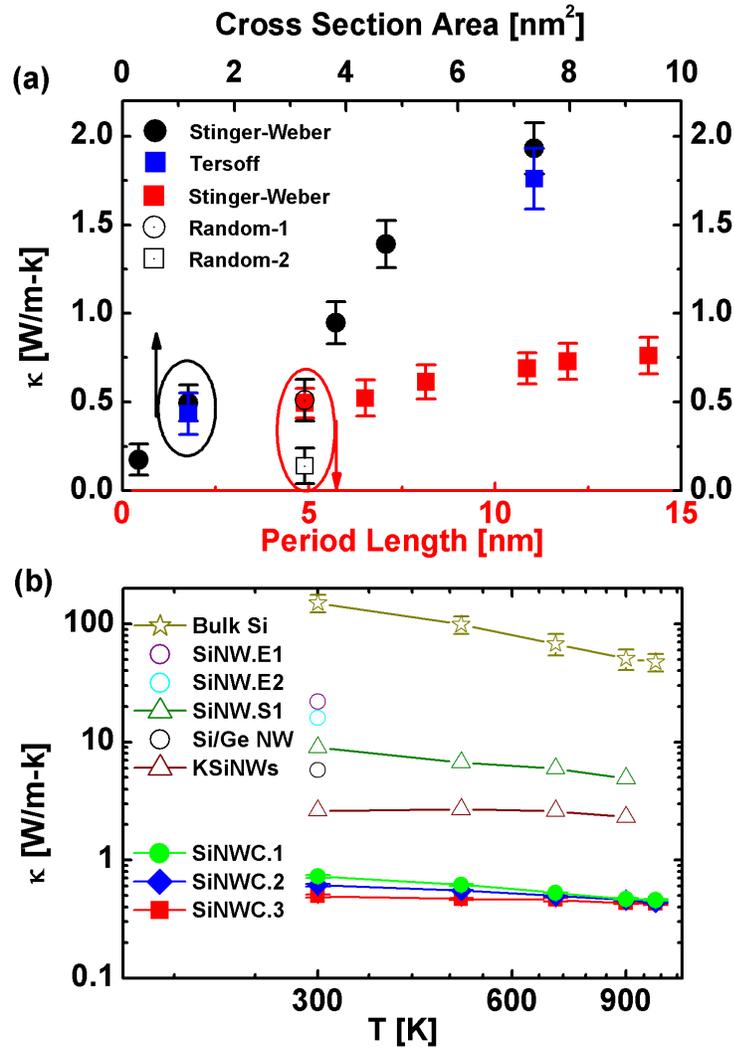

Figure 2. (a) Thermal conductivity of SiNWC versus cross section area of cage bar at 300K, (black dots and blue squares) where the period length is 4.89 nm. The different colors correspond to different potentials. Thermal conductivity of SiNWC versus period length at 300K, (red squares, hollow dot and hollow square) where the cross sectional area is 1.18 nm$^2$. The hollow dot and square correspond to two different random structures. (b) Thermal conductivity of bulk Si, SiNW, Si/Ge NW, kinked SiNW and SiNWC versus the temperature. The cross section area and length or periodic length of SiNWs.E1[31], SiNWs.E2[30], SiNWs.S1[10], Si/Ge NWs[32], KSiNWs[10], SiNWC.1,

SiNWC.2 and SiNWC.3 are 5000 nm$^2$, 17000 nm; 1000 nm$^2$, 150000 nm; 1.21 nm$^2$, 11.9 nm; 5000 nm$^2$, 100-150 nm; 1.21nm$^2$, 19 nm, 1.18 nm$^2$, 11.9 nm; 1.18 nm$^2$, 8.15 nm and 1.18 nm$^2$, 4.89 nm, respectively. In this figure, E1, E2 and S1 correspond to the experimental and simulation results, respectively. $\kappa_{NWC}$ is the effective thermal conductivity which is calculated by the effective volume of all Si atoms, not the volume of whole simulation cell.

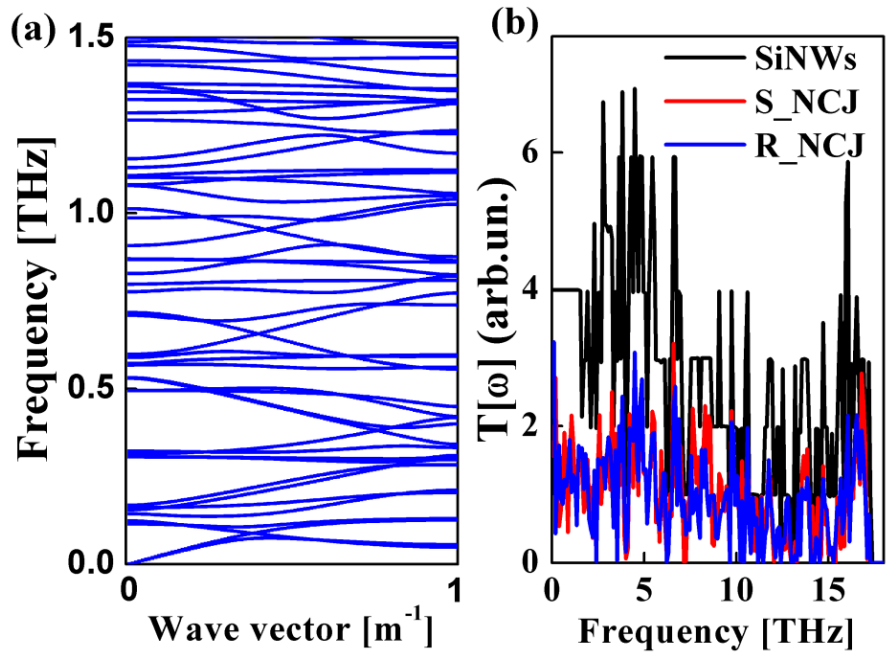

Figure 3. (a) The phonon dispersion relationship of SiNWC from 0~1.6 THz. (Other frequencies are shown in Fig. A4). (b) The phonon transmission coefficient of SiNWs (black line), single symmetric NCJ (red line) and single random NCJ (blue line). (3-dimensional atomistic presentations of these structures are shown in Fig. A6)

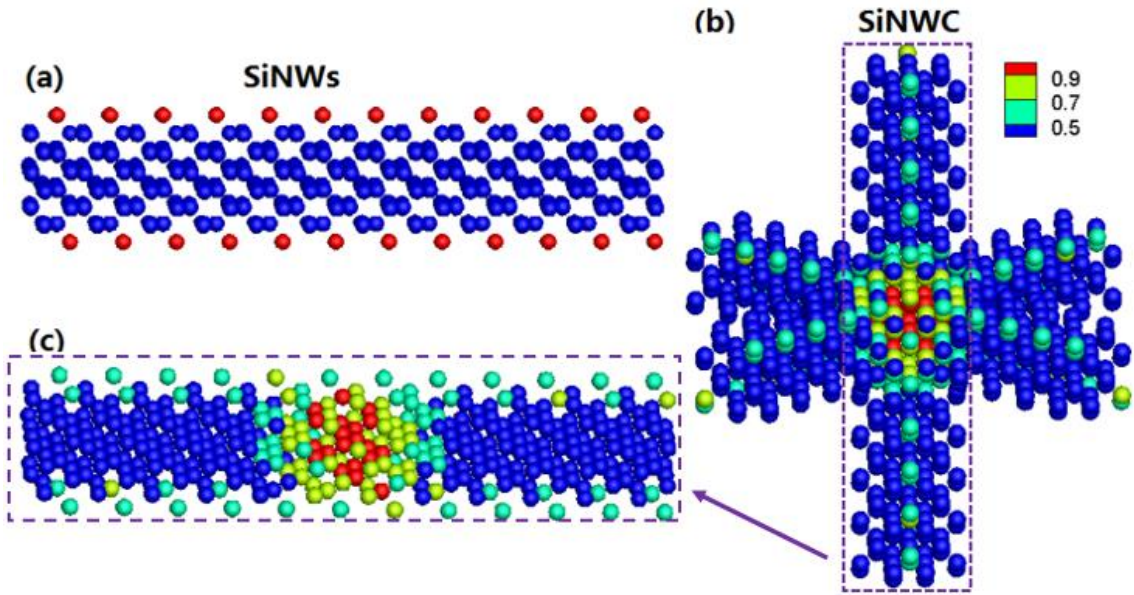

Figure 4. Normalized energy distribution of (a) a SiNW and (b) a SiNWC. The length and cross section area of SiNW are 6.52nm and 1.18 nm$^2$, respectivily, which are the same as the bar in SiNWC. (c) The cage bar picked up from sturcture in (b). The intensity of the energy in all three diagrams are depicted according to the color bar on the top right corner.

## APPENDIX I. MD simulation details

Figure 1(c) shows the simulation cell in MD. The periodic boundary condition is applied in all three directions in simulations. For the force field of covalently bonded Si, we use Stillinger-Weber (SW) potential, which includes both two-body and three-body potential terms. The SW potential has been used widely to study the thermal property of SiNWs and Si bulk material for its best fit to experimental results of the thermal expansion coefficients. The heat flux is defined as:

$$\vec{J}_l(t) = \sum_i \vec{v}_i \varepsilon_i + \frac{1}{2} \sum_{ij(i \neq j)} \vec{r}_{ij}(\vec{F}_{ij} \cdot \vec{v}_{ij}) + \sum_{ijk} \vec{r}_{ij}(\vec{F}_j(ijk) \cdot \vec{v}_j) \tag{A1}$$

where $\vec{F}_{ij}$ and $\vec{F}_j(ijk)$ denote the two-body and three-body forces, respectively. The thermal conductivity is obtained from the Green-Kubo formula:

$$\kappa = \frac{1}{3Vk_B T^2} \int_0^\infty \langle \vec{J}(0) \cdot \vec{J}(\tau) \rangle d\tau \tag{A2}$$

where V is the volume of all the Si atoms (solid part of SiNWC excluding the void), $k_B$ is the Boltzmann constant, $\vec{J}(\tau)$ is the heat flux, T is the temperature and the angular bracket denotes an ensemble average.

The velocity Verlet algorithm is employed in integrating equations of motion, and the time step is 0.25 fs. Initially, the system is equilibrated under the canonical ensemble (NVT) with the Langevin heat reservoir at the target temperature for 0.1 ns, followed by relaxation under a microcanonical ensemble (NVE) for 0.12 ns. The heat current is then recorded with NVE ensemble for 1.2 ns to calculate the thermal conductivity. The value of thermal conductivity is the mean value of twelve simulations with different initial conditions.

Thermal conductivity is derived from the Green-Kubo formula (equation (1)). Fig. A2(a) shows a typical normalized heat current autocorrelation function (HCACF), which is used in Green-Kubo formula to calculate thermal conductivity of SiNWC. The size of these SiNWC is $7.37\times4.89$ nm$^3$ (CSA×PL), $1.18\times10.86$ nm$^3$ and $0.29\times4.89$ nm$^3$, respectively; and the temperature is 300 K. Due to the short relaxation time, the heat current autocorrelation curve decays rapidly at few picoseconds. Then followed by a slower decay to zero within 10 ps approximately. It means the noise is comparable to the signal. Fig. A2(b) shows the thermal conductivity which is an integration of HCACF. The thermal conductivity converges to 1.93, 0.69 and 0.173 Wm$^{-1}$K$^{-1}$, respectively.

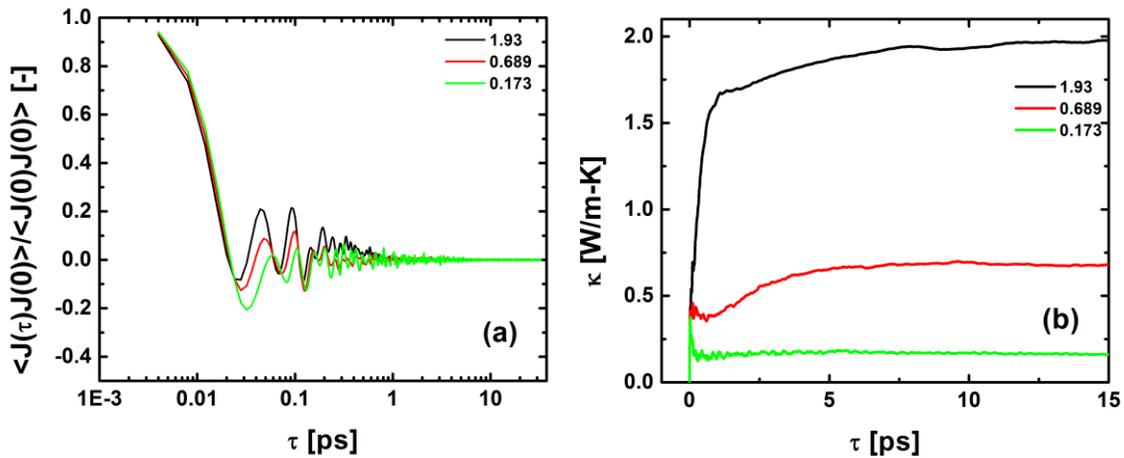

**Figure A1.** (a) Normalized heat current autocorrelation function versus time $\tau$ for SiNWC with different size at 300 K. The cross section area and period length of S1, S2 and S3 are $7.37\times4.89$ nm$^3$, $1.18\times10.86$ nm$^3$ and $0.29\times4.89$ nm$^3$, respectively. This figure shows heat flux correlation rapidly decays to zero in 10 ps. (b) Thermal conductivity calculated by integrating the correlation function in (a) versus time $\tau$. The curve of thermal conductivity converges beyond 10 ps which consistent with the decay of heat current autocorrelation in (a).

For the convenience of comparing with the measurement results, we also present the thermal conductivity calculated by adopting in Eq. (A2) the entire volume (including the void) of SiNWC. The obtained values of thermal conductivity are 0.078, 1.20 and 0.038 $Wm^{-1}K^{-1}$ corresponding to PL=4.89 nm, CSA=1.18 $nm^2$; PL=4.89 nm, CSA=7.37 $nm^2$ and PL=8.85 nm, CSA=1.18 $nm^2$, respectively. It's worth noting that for SiNWC (PL=4.89 nm, CSA=0.29 $nm^2$), the thermal conductivity is 0.0095 $Wm^{-1}K^{-1}$, which is only 0.06‰ of the bulk thermal conductivity of silicon.

## APPENDIX II. Finite size effect in simulations

When using Green Kubo formula to calculate thermal conductivity, the finite size effect could arise if the simulation cell is not sufficiently large. As shown in Figure A1, we calculate thermal conductivity of SiNWC which have different sizes by EMD method at room temperature. We fix period length and cross sectional area as 4.89 nm 1.18 nm$^2$, respectively. The values of thermal conductivity change little when the side length of simulation cell is larger than 4.89 nm. It shows that our simulation cell is large enough to overcome the finite size effect on calculating thermal conductivity. In all of the simulations of SiNWC, we use 9.78 nm as the side length of simulation cell and 4.89 nm as period length.

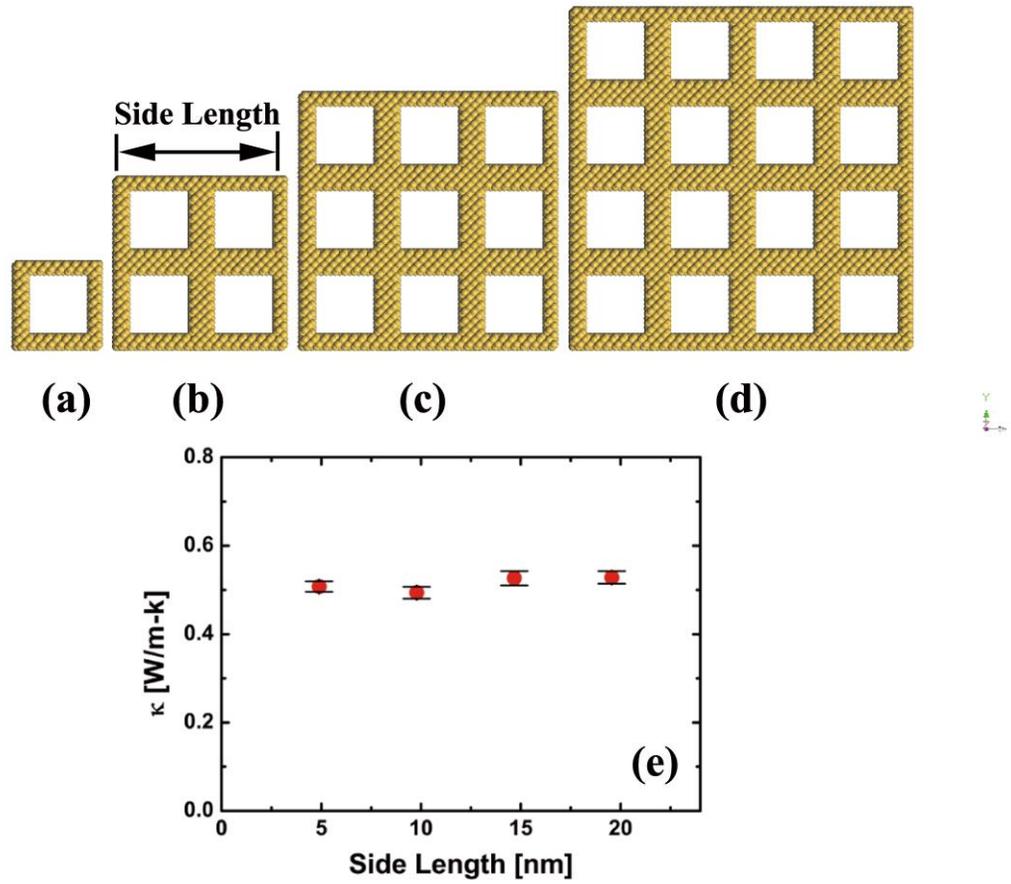

**Figure A2.** (a)-(d) SiNWC (CSA=1.18 nm$^2$, PL=4.89 nm) simulation cell with different side length as 4.89, 9.78, 14.67 and 19.56 nm. (e) The thermal conductivity of SiNWC versus side length of simulation

# APPENDIX III. Schematics of the two random supercell structures

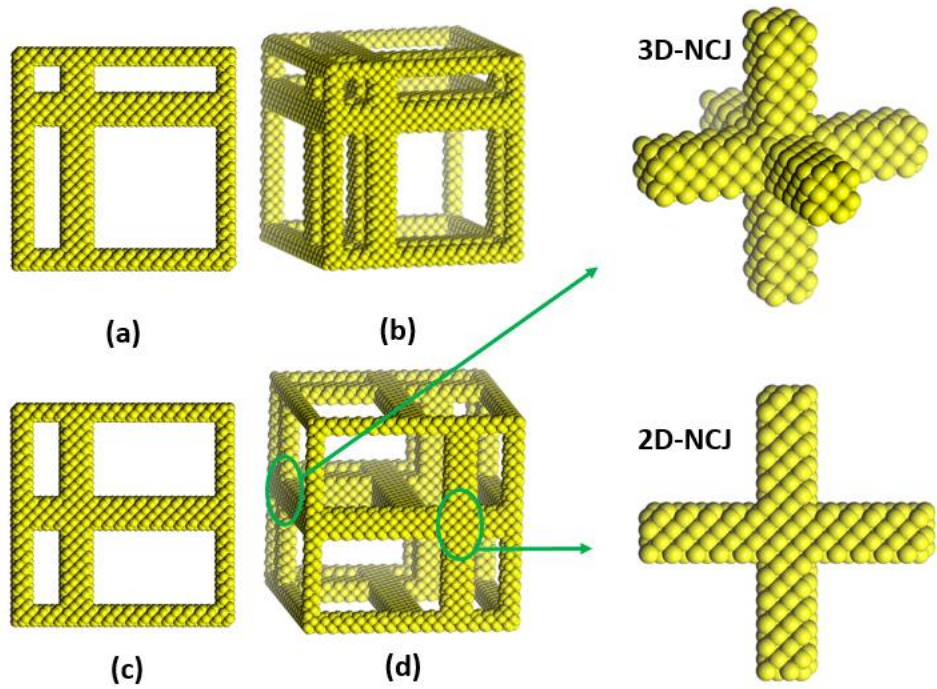

**Figure A3.** (a) (b) Two different views of the first random supercell structure constructed by 3D-NCJs. The number of the 3D-NCJs in this supercell is 8. (c) (d) Two different views of the second random supercell structure constructed by 2D-NCJs and 3D-NCJs. The number of the 2D-NCJs, 3D-NCJs in a supercell are 8 and 3, respectively.

# APPENDIX IV. The phonon dispersion relations of SiNWs and SiNWC by Stillinger-Weber and Tersoff potential.

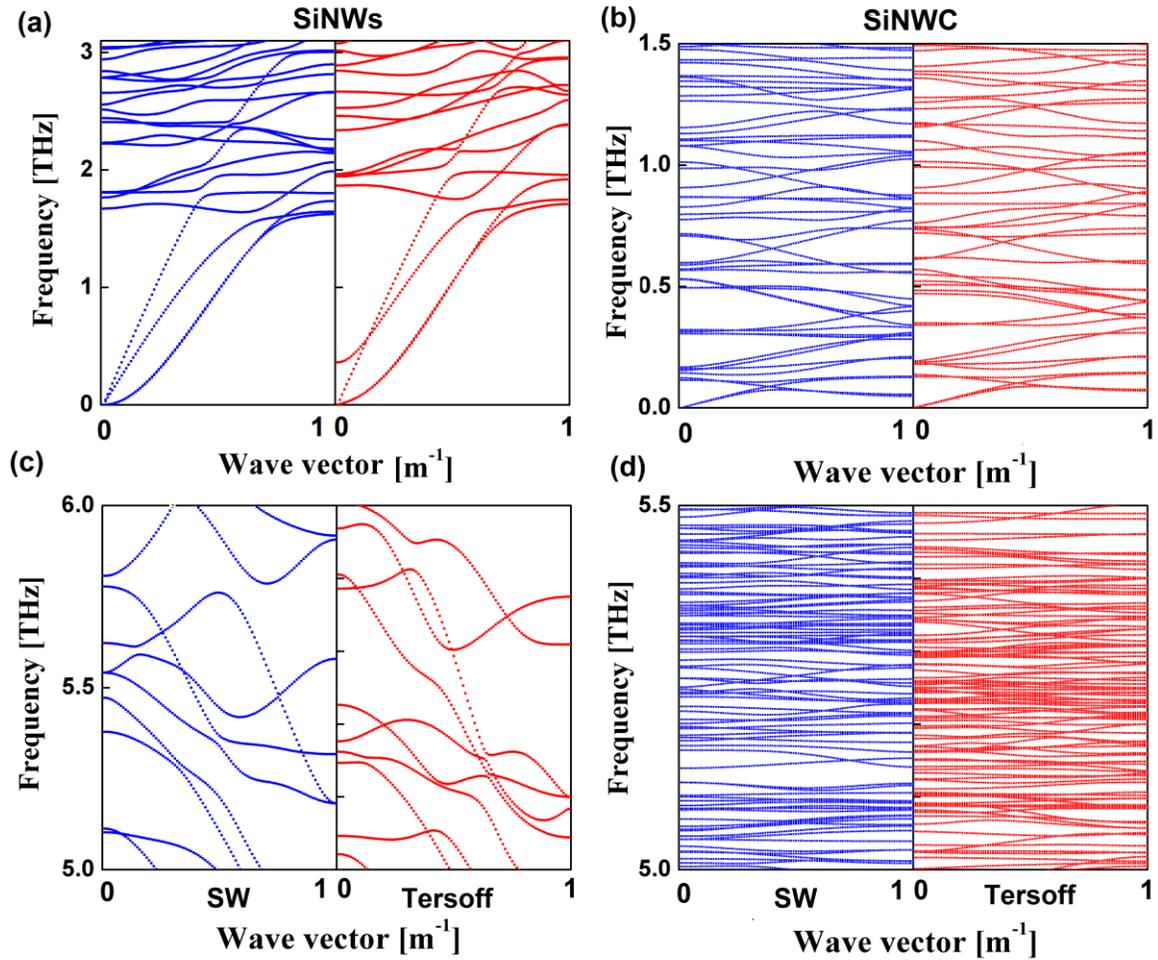

**Figure A4** The phonon dispersion relation of SiNWs by Stillinger-Weber potential (the blue line) and Tersoff potential (the red line) from 0~3THz (a) and 5~6 THz (c). And the phonon dispersion relationship of SiNWC by Stillinger-Weber potential (the blue line) and Tersoff potential (the red line) from 0~1.5THz (b) and 5~5.5 THz (d).

# APPENDIX V. The calculation details of phonon relaxation time

The methodology named normal mode analysis (NMA) can also be found in Henry and Chen's work.[37] Based on LD calculation, the atomic trajectories generated by MD simulations is transformed to normal mode coordinates, which can be expressed as a sum over the displacements of the atoms (labelled as the $j$-th atom in the $l$-th unit cell) in a system as

$$Q(\mathbf{k}, v, t) = \frac{1}{N^{1/2}} \sum_{j,l} m_j^{1/2} \exp(-i\mathbf{k} \cdot \mathbf{r}_{j,0}^l) \cdot \mathbf{e}_j^*(\mathbf{k}, v) \cdot \mathbf{u}_j^l(t), \quad (A3)$$

where $N$ is the number of unit cells in the crystal, $m_j$ is the mass of the atom $j$, $\mathbf{k}$ is the wave vector, $v$ corresponds to the mode polarization, $\mathbf{e}_j^*$ denotes the complex conjugate of eigenvector obtained from lattice dynamics, $\mathbf{r}_{j,0}^l$ and $\mathbf{u}_j^l$ are the equilibrium position and relative displacement from equilibrium position of atom $j$ in unit cell $l$, respectively.

Under the harmonic approximation, the normal mode energy of a classical system can be written as

$$E(\mathbf{k}, v, t) = \frac{\omega^2(\mathbf{k}, v) Q(\mathbf{k}, v, t) \cdot Q^*(\mathbf{k}, v, t)}{2} + \frac{\dot{Q}(\mathbf{k}, v, t) \cdot \dot{Q}^*(\mathbf{k}, v, t)}{2}, \quad (A4)$$

where the first term corresponds to the potential energy and the second term to the kinetic energy.

The phonon mode energy autocorrelation describes the temporal amplitude attenuation, and the frequency of the mode can be identified via the Fourier transform. Then, by fitting the peaks of the normalized autocorrelation with an exponential function, one can obtain the decay time constant, namely the relaxation time of corresponding phonon mode, which can be written as

$$\tau(\mathbf{k}, v) = \frac{\int_0^\infty \langle \delta E(\mathbf{k}, v, 0) \cdot \delta E(\mathbf{k}, v, t) \rangle \mathrm{d}t}{\langle \delta E^2(\mathbf{k}, v, 0) \rangle}, \quad (A5)$$

where the angular bracket denotes an ensemble average.

In this work, we calculate the relaxation time by using NMA, details are described below. In MD simulation and LD calculation, we use the Stillinger-Weber potential to describe the interaction between Si atoms, and the periodic boundary conditions are applied in all three dimensions. To record the displacement and velocity, the simulations are conducted at the temperature of 300K for 12ns and 8ns, time steps are chosen as 1fs and 0.25fs for bulk silicon and SiNWC, respectively. In addition, twelve independent simulations with different initial conditions were conducted to get better average.

The phonon relaxation time of SiNWs is from the work of Martin et al.[4] Because in our simulation the roughness is small, so we choose the SiNWs with the roughness rms (root-mean-square) value equals to 1nm. As shown in Fig. A5, though the cross section of the chosen SiNWs is equal to a circle of 115 nm diameter, the relaxation time is on the order of the SiNWC, especially the low frequency phonons. While the one order reduction of thermal conductivity from bulk silicon to SiNWs come from surface scattering and surface scatterings and length confinements, so the relaxation time has a obvious decline. All of these evidence that the reduction of the thermal conductivity from SiNWs to SiNWC should not come from the surface scatterings, but from nano-cross-junction effect.

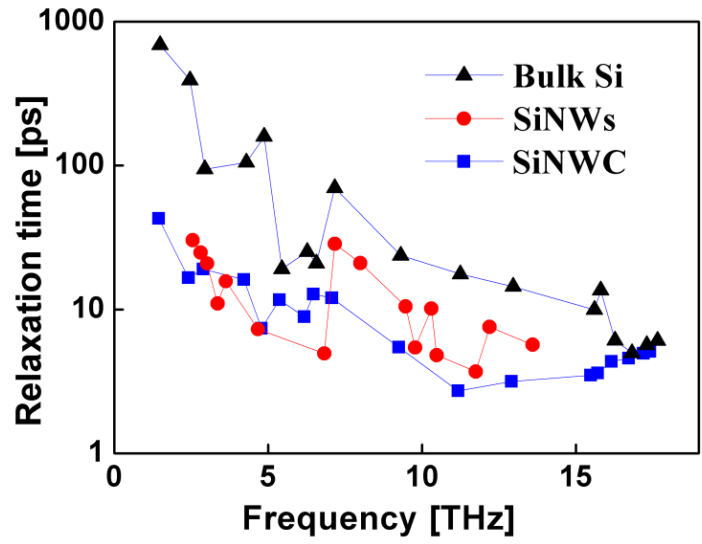

**Figure A5** The relaxation time of bulk silicon (black dot), SiNWs (red dot) and SiNWC (blue dot)

# APPENDIX VI The calculation details of atomistic Green's function method

The targeted system contains three coupled subsystems: two semi-infinite leads connected through the scattering region according to scattering theory. The heat flux flowing in along the system axis write

$$J = \int_{BZ} \hbar \omega_{\mathbf{k}} v_{g,\mathbf{k}_z} (n_L - n_R) t_{\mathbf{k}} \frac{d^3 \mathbf{k}}{(2\pi)^3} \qquad (A6)$$

Where $\hbar \omega_{\mathbf{k}}$ is the energy quantum of the phonon mode $\mathbf{k}$, $v_{g,\mathbf{k}_z}$ is the phonon group velocity of the phonon mode $\mathbf{k}_z$, $n_{L,R}$ is the phonon number on the left and right reservoir following the Bose-Einstein distribution $n = (\exp(\hbar \omega / k_B T) - 1)^{-1}$, $t_{\mathbf{k}}$ is the transmission probability of the phonon mode $\mathbf{k}$. The integration goes through all the phonon modes in the irreducible Brillouin Zone (BZ). In the linear regime, the phonon population undergoes small perturbations and thus the thermal conductance writes

$$G = J / \Delta T = \int_{BZ} \hbar \omega_{\mathbf{k}} v_{g,\mathbf{k}_z} \frac{\partial n}{\partial T} t_{\mathbf{k}} \frac{d^3 \mathbf{k}}{(2\pi)^3} \qquad (A7)$$

We note that $d^3 \mathbf{k} = dk_x dk_y dk_z$ and $v_{g,\mathbf{k}_z} dk_z = \partial \omega / \partial k_z dk_z = d\omega$. The Eq. (A5) reduces to

$$G = J / \Delta T = \int_{BZ} \hbar \omega_{\mathbf{k}} \frac{\partial}{\partial T} (e^{\frac{\hbar w}{k_B T}} - 1)^{-1} [t_{\omega} dk_x dk_y] \frac{d\omega}{(2\pi)^3} \qquad (A8)$$

Hence we identify the spectral phonon transmission function $\Xi(\omega) = t_{\omega} g(\omega)$ where $g(\omega) = dk_x dk_y$ is the projected phonon density of states in the periodic directions of the system.

We probe the spectral phonon transmission function $\Xi(\omega)$ by atomistic green's function (AGF) and the thermal conductance can be obtained by following the Landauer formula:

$$G = \int_0^{\omega_{max}} \Xi(\omega) \frac{\partial}{\partial T} (e^{\frac{\hbar\omega}{k_B T}} - 1)^{-1} \hbar\omega \frac{d\omega}{2\pi} \quad (A9)$$

Where $\hbar\omega_k$ and $\omega_{max}$ are the energy and the maximum frequency in the system. T refers to the mean temperature of the system, $k_B$ and $\hbar$ represent the Boltzmann and the reduced Planck constants, respectively. The transmission $\Xi(\omega)$ is obtained from a nonequilibrium Green's function approach as $Tr[\Gamma_L G_s \Gamma_R G_s^+]$. The advanced and retarded Green functions $G_s^+$ and $G_s$ can be deduced from:

$$G_s = [(\omega + i\Delta)^2 I - K_s - \Sigma_L - \Sigma_R]^{-1} \quad (A10)$$

Where $\Delta$ is an infinitesimal imaginary part that maintains the causality of the Green's function and $\Sigma_L = K_{ab} g_L K_{ab}^+$, $\Sigma_R = K_{ab} g_R K_{ab}^+$ are the self-energies of the left and right leads, the "+" exponent indicating the Hermitian conjugation. Finally, $g_L$ and $g_R$ refer to the surface Green's functions of the left and right leads, while $K_s$ and $K_{ab}$ are the force constant matrices derived from the potential, for the scattering region and between neighboring atoms in the leads, respectively. The expression of the transmission also includes $\Gamma_L = i(\Sigma_L - \Sigma_L^+)$ and $\Gamma_R = i(\Sigma_R - \Sigma_R^+)$.

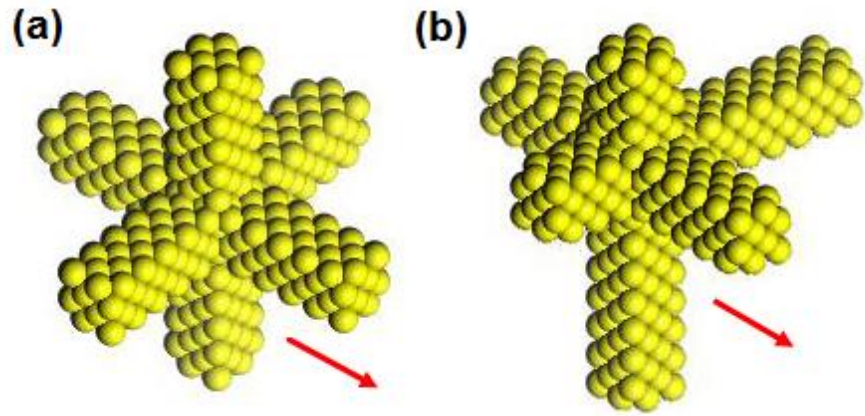

**Figure A6** (a) The symmetric nano-cross-junction structure. (b) The random nano-cross-junction structure. The transmission coefficient is calculated in the direction shown by the red arrows.

# APPENDIX VII. The calculation details of participation ratio and energy distribution.

The localization of each eigen-mode, $\lambda$, can be quantitatively characterized by participation ratio, $P_\lambda$, which is defined as

$$P_\lambda^{-1} = N \sum_i \left( \sum_\alpha \varepsilon_{i\alpha,\lambda}^* \varepsilon_{i\alpha,\lambda} \right)^2 \tag{A11}$$

where N is the total number of atoms, $\varepsilon_{i\alpha,\lambda}$ is the αth eigen-vector component of $\lambda$ for atom i. The participation ratio measures the fraction of atoms participating in a given mode. The value of $P_\lambda$ corresponds to a localized mode with O(1/N) and a delocalized mode with O(1).

The spatial distribution of energy is calculated as

$$E_i = \sum_\omega \sum_\lambda \sum_\alpha (n+1/2)\, \hbar\omega \varepsilon_{i\alpha,\lambda}^* \varepsilon_{i\alpha,\lambda} \delta(\omega-\omega_\lambda) \tag{A12}$$

where *n* is the phonon occupation number given by the Bose-Einstein distribution.

# APPENDIX VIII. The participation ratio calculated by Stillinger-Weber and Tersoff potential

The phonon eigen-frequencies and eigen-vectors are obtained by lattice dynamics (LD). When using the general utility lattice program (GULP), the periodic boundary condition and free boundary condition were applied in the longitudinal direction and lateral direction of both SiNWs and SiNWC, respectively. The boundary conditions are consistent with those in our MD simulations so that the phonon eigen-modes obtained by LD becomes consistent with those existing in MD.

To show the phonon localization in bulk SiNWC. In Fig. A7a, we compare the participation ratio (P) of each eigen-mode for infinite length SiNWs (blue dots) with those for the SiNWC (red dots). The period length and cross section area of SiNWs are 6.52 nm and 1.18 $nm^2$, respectively, which are the same as the size of cage bar of the SiNWC. It shows a clear reduction of participation ratio in SiNWC for both low and high frequency phonons. For the SiNWs, most of the P values are larger than 0.5, which indicates delocalized phonon modes. While, for the SiNWC, most of the P values are less than 0.5, which means these modes are likely to be localized.

To quantitatively analyze the phonon localization, we define the localization ratio (LR) as the number of localized modes, whose P value is less than 0.5, divided by the total number of modes. According to the definition, a larger LR value corresponding to more localized phonon modes. Also shown in Fig. A7a, the LR value of SiNWC is 78.4%, which is more than two times larger than that of SiNWs (33.1%). It means that the introduction of cross junction drastically makes more phonon modes localized, which gives rise to the ultralow and temperature-independent thermal conductivity.

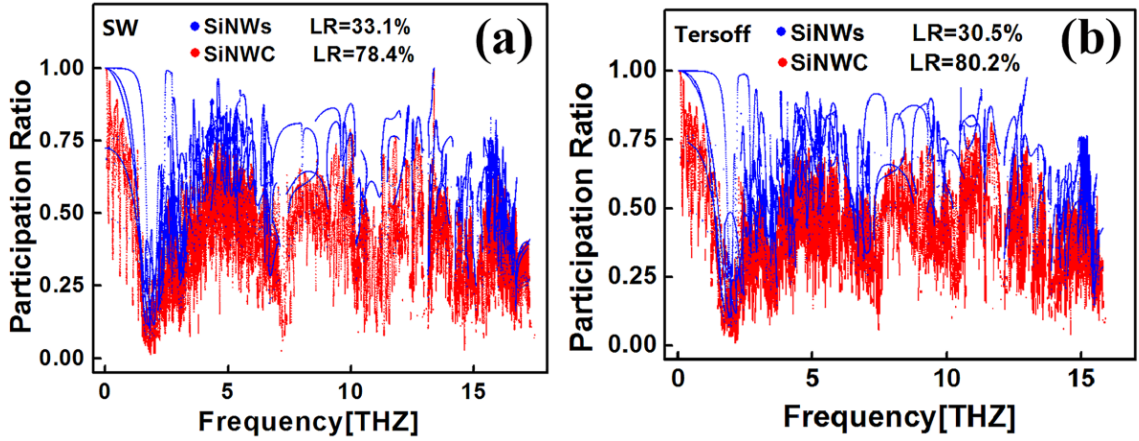

**Figure A7.** Participation ratio of each eigen-mode for SiNWs (blue dots) and SiNWC (red dots) calculated by (a) Stillinger-Weber and (b) Tersoff potential. LR values are also shown according to the definition.

# APPENDIX IX. The normalized energy distribution of SiNWs and SiNWC calculated by Tersoff potential

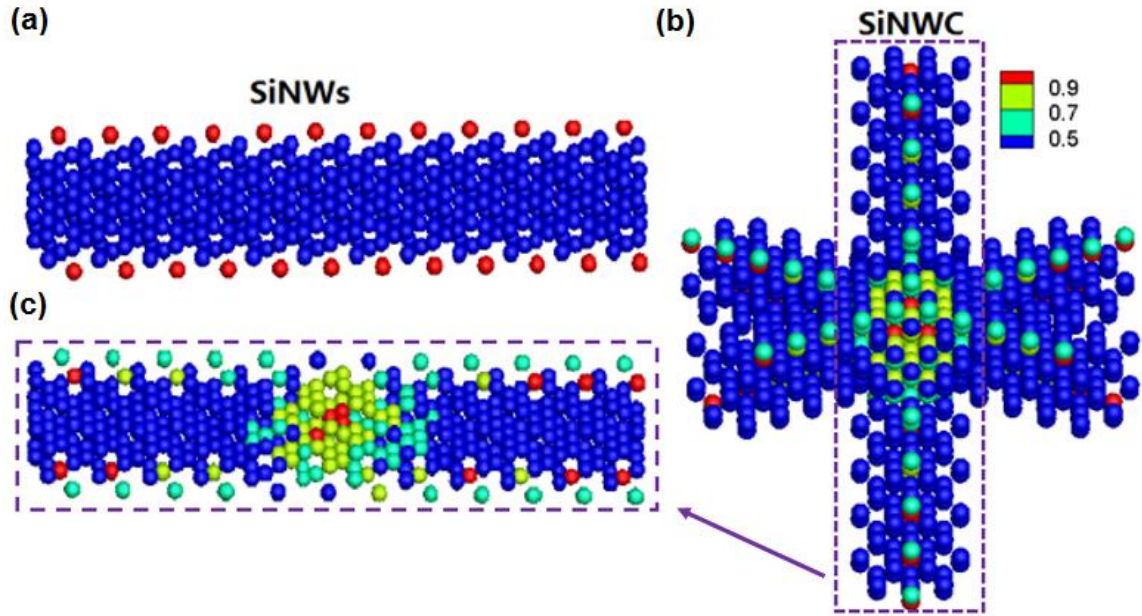

Figure A8. Normalized energy distribution of (a) a SiNW and (b) a SiNWC calculated by Tersoff potential. The length and cross section area of SiNW are 6.52nm and 1.18 nm$^2$, respectively, which are the same as the bar in SiNWC. (c) The cage bar picked up from structure in (b). The intensity of the energy in all three diagrams are depicted according to the color bar on the top right corner.

# APPENDIX X. A comparison with the work of Hussein et al.

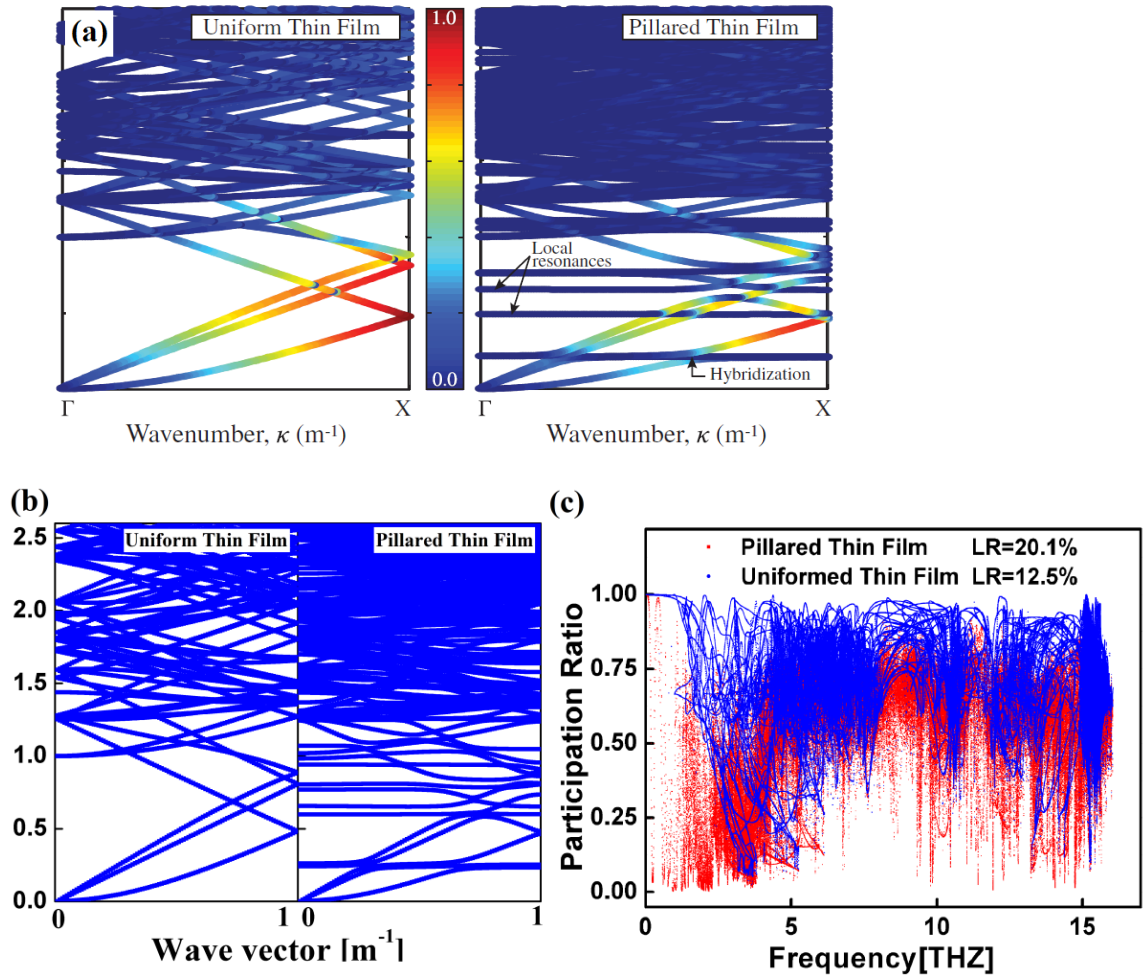

Figure A9 (a) The phonon dispersion relations of UTF and PTF calculated by Hussein et al. (b) The phonon dispersion relations of UTF and PTF calculated by us. (c) the participation ratio of UTF and PTF calculated by us, and LR values are also shown according to the definition.

According to the data of Hussein et al.,[18] we construct the same uniform thin film (UTF) and pillared thin film (PTF) structure. Due to our Tersoff potential, the side length a= 0.543 nm which is slightly different from Hussein's (a= 0.54 nm). The phonon dispersion relations of UTF and PTF we calculated (shown in Fig. A9b) by Tersoff potential is almost the same as the work of Hussein et al (shown in Fig. A9a).

Then, we calculated the participation ratio (PR) of UTF and PTF (shown in Fig. A9b). The PR of PTF shows a clear reduction in low frequency phonons, while in other frequencies, only few modes show a reduction in PR. A more detailed calculation shows that, in Hussein's work, the localization ratio value increases from 12.5% (UTF) to 20.1% (PTF). While in our structures, (as shown in Fig. A7) the LR value hugely increases from 30.5% (SiNWs) to 80.2% (SiNWC).